\documentclass[sigconf]{acmart}
\setcopyright{rightsretained}

\usepackage{amsmath,amssymb,amsfonts}
\usepackage{graphicx}
\usepackage{textcomp}
\usepackage{diagbox}
\usepackage{booktabs}
\usepackage{colortbl}
\usepackage{multirow}
\usepackage{tabularx}
\usepackage{soul}
\usepackage{wrapfig}
\usepackage[caption=false]{subfig}
\usepackage[font=footnotesize,labelfont=bf]{caption}
\usepackage[normalem]{ulem}
\usepackage{algorithm}
\usepackage[noend]{algpseudocode}
\algtext*{EndWhile}
\algtext*{EndIf}
\algtext*{EndFor}
\algnewcommand\algorithmicforeach{\textbf{for each}}
\algdef{S}[FOR]{ForEach}[1]{\algorithmicforeach\ #1\ \algorithmicdo}

\usepackage{xcolor}
\definecolor{myblue}{HTML}{93CDDD}
\definecolor{mygreen}{HTML}{17891B}
\definecolor{myyellow}{HTML}{DCC60F}
\definecolor{myorange}{HTML}{EB801B}
\usepackage{enumitem}
\usepackage{footnote}  
\usepackage{pifont}

\AtBeginDocument{%
  }

\copyrightyear{2025}
\acmYear{2025}
\setcopyright{rightsretained}
\acmConference[GLSVLSI '25]{Great Lakes Symposium on VLSI 2025}{June 30-July 2, 2025}{New Orleans, LA, USA}
\acmBooktitle{Great Lakes Symposium on VLSI 2025 (GLSVLSI '25), June 30-July 2, 2025, New Orleans, LA, USA}
\acmDOI{10.1145/3716368.3735248}
\acmISBN{979-8-4007-1496-2/2025/06}

\begin{document}


\title{SAFE-SiP: \underline{S}ecure \underline{A}uthentication \underline{F}ram\underline{e}work for \underline{S}ystem-\underline{i}n-\underline{P}ackage Using Multi-party Computation}

\author{Ishraq Tashdid}
\affiliation{
  \department{Department of Electrical and Computer Engineering}
  \institution{University of Central Florida}
  \city{Orlando}
  \state{FL}
  \country{USA}
}
\email{ishraq.tashdid@ucf.edu}

\author{Tasnuva Farheen}
\affiliation{
  \department{Department of Computer Science and Engineering}
  \institution{Louisiana State University}
  \city{Baton Rouge}
  \state{LA}
  \country{USA}
}
\email{tfarheen@lsu.edu}

\author{Sazadur Rahman}
\affiliation{
  \department{Department of Electrical and Computer Engineering}
  \institution{University of Central Florida}
  \city{Orlando}
  \state{FL}
  \country{USA}
}
\email{mohammad.rahman@ucf.edu}

\begin{abstract}

The emergence of chiplet-based heterogeneous integration revolutionizes semiconductor, AI, and high-performance computing systems by enabling modular design and enhanced scalability. However, the post-fabrication assembly of chiplets from multiple vendors introduces a complex supply chain, raising critical security concerns such as counterfeiting, overproduction, and unauthorized access. Existing solutions rely on dedicated security chiplets or modifications to the timing flow that inherently assumes a trusted SiP integrator, exposing chiplet signatures to other vendors and introducing additional attack surfaces. This work addresses these vulnerabilities by leveraging Multi-party Computation (MPC), which ensures zero-trust authentication without revealing sensitive information to any party. We introduce \emph{SAFE-SiP}, a scalable authentication framework that garbles chiplet signatures and employs MPC for integrity verification, preventing unauthorized access and adversarial inference. \emph{SAFE-SiP} eliminates the need for a dedicated security chiplet while ensuring authentication remains secure, even in untrusted integration environments. We evaluated \emph{SAFE-SiP} across five RISC-V-based SiPs. Our experimental results shows that, \emph{SAFE-SiP} achieves minimal power overhead, incurs an average area overhead of only $3.05\%$, and maintains a computational complexity of $2^{192}$, providing a highly efficient and scalable security solution.

\end{abstract}

\maketitle

\section{Introduction}
\begin{figure}[t] 
    \centering
    \includegraphics[width=0.5\textwidth]{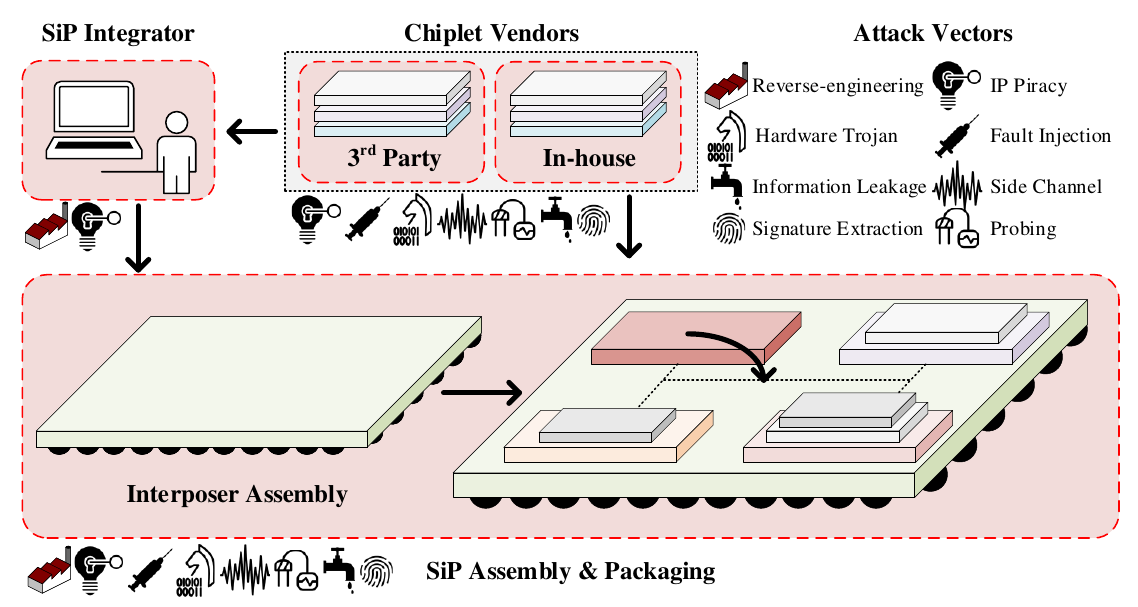}
    \caption{SiP design flow and life-cycle using third-party and in-house chiplets. Untrusted and trusted parties are marked by red and green boxes respectively.}
    \label{fig:threatmodel} 
\end{figure}

\noindent The shift from monolithic System-on-Chip (SoC) architectures to System-in-Package (SiP) based heterogeneous integration (HI) is crucial for sustaining yield in advanced nodes, co-integrating diverse technologies, and enabling 3D stacking for improved performance and efficiency~\cite{3dic_book_pavlidis, 3d_soictm}. By modularizing functionalities into chiplets, SiP enhances design flexibility at the cost of challenges in co-design, reliability, and security~\cite{ieee1838_casestudy, 2.5_3D_hetero}. Unlike traditional SoCs, SiP requires post-fabrication assembly of multi-vendor chiplets, often in untrusted environments, making it vulnerable to counterfeiting, overproduction, spoofing and unauthorized modifications~\cite{gate_sip, haque2023shi}. The reliance on untrusted foundries for interposer fabrication and SiP integration further exacerbates risks such as hardware trojan insertion and backdoor attacks~\cite{toshi}. Additionally, counterfeit chiplets infiltrating the supply chain threaten functional integrity and security. As shown in Fig.~\ref{fig:threatmodel}, ensuring chiplet authenticity pre- and post-fabrication is vital to system trustworthiness~\cite{pqc_hi}. Without a scalable authentication mechanism, SiP-based HI remains susceptible to exploitation at supply chain stages.

Recent research highlights the unique security challenges in SiP but also exposes the limitations of existing approaches. Traditional techniques such as logic locking~\cite{haque2023shi,rahman2020defense}, watermarking~\cite{watermark_actiwate, anandakumar2022}, and IC metering~\cite{alkabani2007active} are effective for SoCs but struggle with the multi-vendor complexity of SiP~\cite{ll_seamless, postfab_lle}. Recent initiatives, such as, GATE-SiP~\cite{gate_sip}, PQC-HI~\cite{pqc_hi}, and SECT-HI~\cite{sect_hi} improve security through test access port (TAP) modifications, quantum-resistant authentication, and hardware security modules but introduce significant area overhead. Fabrication-level techniques such as split manufacturing and network-on-interconnect (NoI) obfuscation mitigate risks like cloning and tampering~\cite{splitcore, script_dos}. However, split manufacturing is costly and depends on trusted BEOL foundries, facing yield and scalability issues~\cite{twofold_laser, postfab_lle}, while NoI obfuscation increases design complexity. These limitations underscore the need for scalable, cost-effective security solutions tailored to heterogeneous systems.

Multi-party Computation (MPC) is a cryptographic protocol that enables multiple parties to collaboratively compute a function over their private inputs while ensuring data privacy and correctness, even in zero-trust environments. The SiP assembly fits well with MPC as the integrator can use a function to evaluate the authenticity of the chiplets without knowing the actual signatures from the integrated chiplets. Moreover, garbling circuits offer a robust solution to obfuscate signatures in an HI environment by balancing security resiliency and cost. Recent works like TinyGarble~\cite{tinygarble} and MPCircuits~\cite{mpcircuits} enhance efficiency and scalability in secure computations and multi-party protocols while maintaining low area overhead. Inspired by these, we consider garbling circuits to obscure chiplet signatures and employ secure MPC, safeguarding operational logic even if intercepted. Hence, this paper introduces \emph{SAFE-SiP}, an MPC-based framework that integrates seamlessly with chiplets, adding a universal garbling scheme for authentication.

\noindent\textbf{Contributions.} Our main contributions are summarized below.
\vspace{-0.05in}
\begin{enumerate}[leftmargin=*]
    \item We propose \emph{SAFE-SiP}, a multi-party computation-based authentication framework using garbling circuits to secure chiplet integration in 2.5D/3D SiP assemblies, ensuring authentication, and secure testing while preserving data confidentiality. It is designed for broader compatibility, seamlessly integrating with diverse signature processes and supporting a secure boot mechanism. \textbf{\uline{To the best of our knowledge, this is the first chiplet authentication scheme in a zero-trust threat model.}}

    \item We perform a comprehensive security analysis demonstrating \emph{SAFE-SiP}'s resilience against threats, with garbling and SHA-256 protecting against tampering, spoofing, and replay attacks.
    \item We evaluate \emph{SAFE-SiP} on five RISC-V-based benchmarks, achieving as low as $1.84$\% area overhead while maintaining a computational complexity of \(2^{192}\) for a $64$ bit security parameter.
    \item To encourage community collaboration and industry adoption, we plan to open-source our implementation at \url{https://github.com/IshraqAtUCF/safe_sip} following publication.

\end{enumerate}
\vspace{-0.05in}

\noindent The rest of the paper is organized as follows—Section~\ref{sec:motivation} outlines the threat model, reviews existing authentication methods, and motivates a secure, low-overhead SiP framework. Section~\ref{sec:methodology} presents \emph{SAFE-SiP}, detailing its MPC-based authentication, garbling circuits, and SHA-256 watermark protection. Section~\ref{attack} analyzes security against removal, replay, tampering, and forgery. Section~\ref{sec:result} evaluates in real-world settings, and Section~\ref{sec:conclusion} concludes the paper.
\vspace{-0.05in}
\begin{table}[t]
\centering
\caption{Attack Vectors in SiP Supply Chain.}
\label{tab:attackvectors}
\setlength{\tabcolsep}{1pt} 
\renewcommand{\arraystretch}{1.2} 
\resizebox{0.45\textwidth}{!}{%
\begin{tabular}{ccccc}
\toprule
\diagbox{Scenario}{Entity} 
    & \begin{tabular}[c]{@{}c@{}}SiP \\Integrator\end{tabular}
    & \begin{tabular}[c]{@{}c@{}}3rd Party \\Vendors\end{tabular}
    & \begin{tabular}[c]{@{}c@{}}Interposer \&\\ Packaging Foundry\end{tabular}
    & \begin{tabular}[c]{@{}c@{}}Attack\\Vector\end{tabular} \\
\midrule
Scenario A 
    & Trusted 
    & Untrusted 
    & Untrusted 
    & \begin{tabular}[c]{@{}c@{}}Side-Channel Attack\\Unauthorized Probing\end{tabular} \\
\midrule
Scenario B
    & Untrusted 
    & Trusted 
    & Untrusted 
    & \begin{tabular}[c]{@{}c@{}}Signature Extraction\end{tabular} \\
\midrule
Scenario C 
    & Trusted 
    & Trusted 
    & Untrusted 
    & \begin{tabular}[c]{@{}c@{}}Hardware Trojan\\Overproduction\end{tabular} \\
\bottomrule
\end{tabular}%
}
\end{table}
\section{Background and Motivation} \label{sec:motivation} 
In this section, we discuss the complex SiP supply chain, the threat model, and existing system-level authentication mechanisms.
\vspace{-0.1in}

\subsection{SiP Supply Chain and Threat Model}\label{subsec:threat_model}

\noindent Fig.~\ref{fig:threatmodel} illustrates the multifaceted security threats in SiP-based heterogeneous integration, highlighting vulnerabilities across chiplet vendors, SiP integrators, and foundries under a zero-trust paradigm. Third-party vendors are particularly susceptible to unauthorized probing and side-channel attacks (Scenario A, Table~\ref{tab:attackvectors}), where adversarial chiplets leverage information leakage techniques to extract proprietary signatures and compromise vendor confidentiality. As depicted in Fig.~\ref{fig:threatmodel}, adversarial chiplets can inject malicious modifications during fabrication, embedding covert circuits capable of intercepting authentication processes. The SiP integrator, despite being trusted in some scenarios, may also pose an untrusted environment (Scenario B), where it attempts signature extraction attacks through reverse engineering or unauthorized tampering to reuse or overproduce chiplets. This scenario is exacerbated by the presence of a compromised interposer or packaging foundry, which can facilitate the overproduction of counterfeit chiplets or introduce hardware Trojans to manipulate authentication mechanisms (Scenario C). Fig.~\ref{fig:threatmodel} further illustrates that during SiP assembly and packaging, the interposer can serve as a conduit for various attack vectors, including hardware trojans, fault injection, and probing. The foundry, which plays a critical role in the integration process, poses significant threats by embedding backdoors within the interposer or modifying circuit layouts to facilitate unauthorized surveillance of inter-chiplet communication. These attack surfaces collectively underscore the necessity of a zero-trust authentication model that prevents any entity—whether vendor, integrator, or foundry—from gaining unauthorized access to chiplet authentication data.

\vspace{-6pt}
\subsection{Existing Works and their Drawbacks}\label{subsec:existing_techniques}
Recent research on chiplet security focuses on fabrication-level techniques such as split manufacturing (SM) and NoI obfuscation to mitigate tampering and unauthorized access. SM obscures chiplet interconnections, reducing security risks~\cite{splitcore}, while secure routing disrupts predictable NoI paths to prevent DDoS attacks~\cite{script_dos}. However, reliance on trusted BEOL foundries limits scalability, and SM introduces FEOL-BEOL alignment challenges that may impact yield and functionality~\cite{postfab_lle}. These constraints highlight the need for holistic security solutions in chiplet-based systems. Researchers have explored integrating Chiplet Hardware Security Modules (CHSM) and Chiplet Security Intellectual Property (CSIP) into System-in-Package (SiP) architectures~\cite{toshi}. As shown in Tab.~\ref{tab:intro_comparison}, PQC-HI employs post-quantum cryptography for chiplet authentication and key exchange, protecting against probing and unauthorized data extraction from active interposers~\cite{pqc_hi}. SECT-HI, meanwhile, secures the SiP testing phase by encrypting scan chain outputs and embedding watermarks, ensuring only verified SiPs are deployed~\cite{sect_hi}. However, CHSM and CSIP add design complexity and cost, potentially attracting attackers if these modules become bypass targets, raising concerns about their practical adoption.

\begin{figure}[t] 
    \centering
    \includegraphics[width=0.425\textwidth]{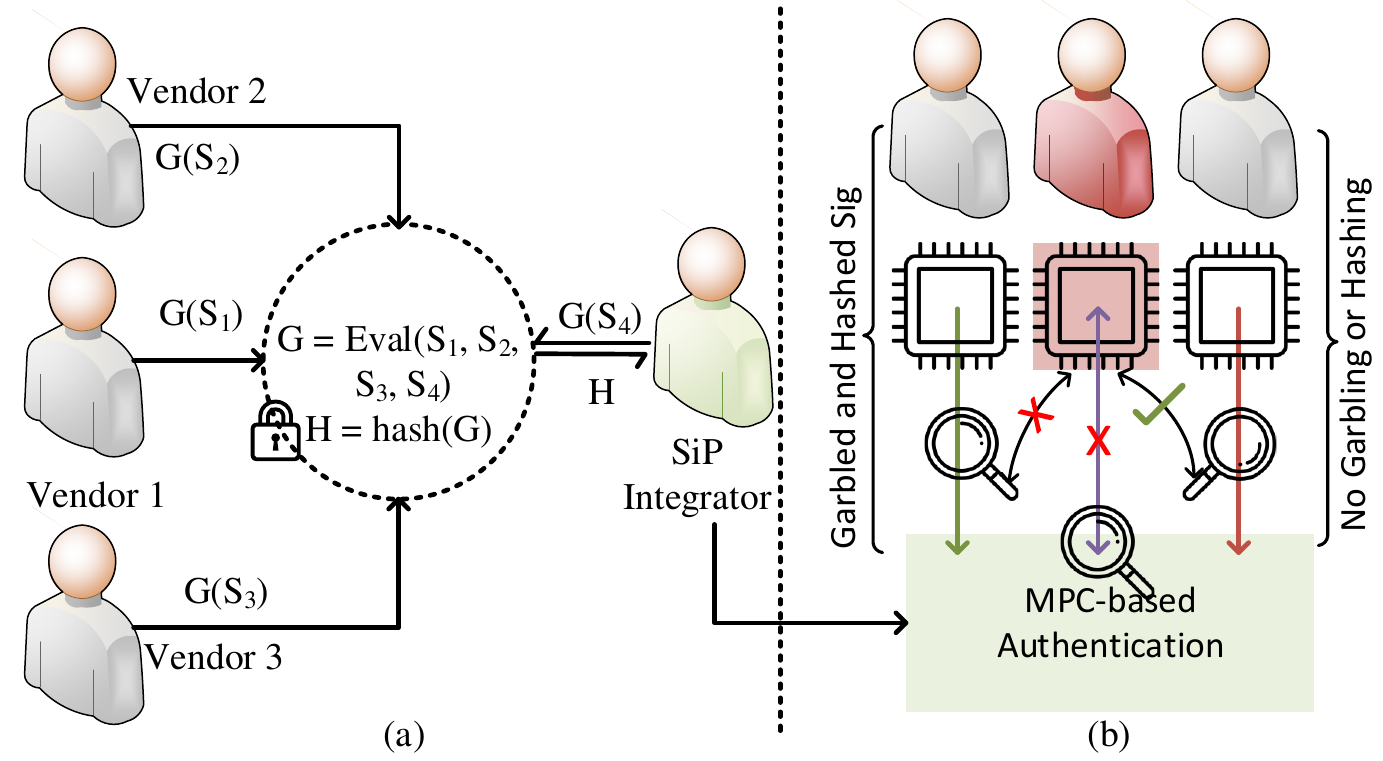}
    \caption{(a) Multi-party Computation scheme and (b) adversarial probing from malicious chiplet or 3rd-party foundry mitigated by \emph{SAFE-SiP}.}
    \label{fig:mpc} 
\end{figure}

\begin{table}[t]
\caption{Comparison with State-of-the-art Solutions (SOTA).}
\centering
\label{tab:intro_comparison}
\resizebox{0.475\textwidth}{!}{
\setlength{\tabcolsep}{1pt} 
\begin{tabular}{ccc} 
\toprule
\textbf{\begin{tabular}[c]{@{}c@{}}Existing \\Technique\end{tabular}} & 
\textbf{Limitations} & 
\textbf{\begin{tabular}[c]{@{}c@{}}\emph{SAFE-SiP} \end{tabular}} \\
\midrule
\textbf{GATE-SiP~\cite{gate_sip}} &  
\begin{tabular}[c]{@{}c@{}}TAP-based authentication; \\ Vulnerable to MITM attacks\end{tabular} &  
\begin{tabular}[c]{@{}c@{}}No modification to the TAP; \\ Unperturbed from MITM attacks\end{tabular} \\
\midrule
\textbf{PQC-HI~\cite{pqc_hi}} &  
\begin{tabular}[c]{@{}c@{}}High computational overhead; \\ Susceptible to probing attacks\end{tabular} &  
\begin{tabular}[c]{@{}c@{}}Lightweight authentication with; \\ Strong signature obfuscation\end{tabular} \\
\midrule
\textbf{SECT-HI~\cite{sect_hi}} &  
\begin{tabular}[c]{@{}c@{}}Limited to test encryption only; \\ Restricts vendor security control\end{tabular} &  
\begin{tabular}[c]{@{}c@{}}Considers vendor's stake in SiP; \\ Security for both vendor \& integrator\end{tabular} \\
\midrule
\textbf{\begin{tabular}[c]{@{}c@{}}Know Time \\to Die \cite{deric2022know} \end{tabular}} &  
\begin{tabular}[c]{@{}c@{}}Prone to challenge-response pair attacks; \\ Sensitive to environmental variations\end{tabular} &  
\begin{tabular}[c]{@{}c@{}}Environment-agnostic; \\ Cryptographic authentication\end{tabular} \\
\bottomrule
\end{tabular}}
\begin{minipage}{\linewidth}
\vspace{1pt} 
\footnotesize \textit{* Unlike SOTA except~\cite{deric2022know}, \emph{SAFE-SiP} operates without dedicated security chiplet.}
\end{minipage}

\end{table}

\vspace{-0.1in}

\section{SAFE-SiP Methodology} \label{sec:methodology}
In this section, we provide an overview of the \emph{SAFE-SiP}, discuss the detailed framework, and secure-boot driven communication flow.
\vspace{-0.1in}

\subsection{Security Objectives: Zero-trust Threat Model}\label{subsec:sec_obj}

\noindent A zero-trust threat model follows `trust but verify' paradigm, where no entity within the SiP supply chain can be inherently trusted. Based on the threat model discussed in Sec.~\ref{subsec:threat_model} we propose the following two security objectives for chiplet authentication.

\vspace{0.1in}
\noindent\underline{\textbf{(SecObj1)}} The interposer, facilitating communication between heterogeneous chiplets, is a critical vulnerability point due to the lack of a secure perimeter, exposing data to probing, spoofing, and man-in-the-middle attacks~\cite{deric2022know}. Securing this layer is essential to ensure system integrity, especially when chiplets from untrusted sources interact without guaranteed security features. Hence, authentication must occur without exposing sensitive data, necessitating cryptographic techniques, such as, garbling circuits. Garbled circuit is a cryptographic protocol enabling secure MPC, where two parties jointly evaluate a function on private inputs without revealing them~\cite{gc_fairplay, gc_secure_auction}. As shown in Fig.~\ref{fig:mpc}, vendors encode signatures into garbled values ($G(S_{i})$), which the SiP integrator processes using a trusted chiplet without direct access to secret signatures~\cite{gc_fkb, gc_yao}. This enforces encrypted authentication, preventing both integrators and foundries from accessing vendor signatures. Thus mitigating side-channel attacks, probing (Scenario A, Table~\ref{tab:attackvectors}), snooping, malicious modifications, and other attack scenarios (Scenario B, C). 

\begin{figure}[t] 
    \centering
    \includegraphics[width=0.5\textwidth]{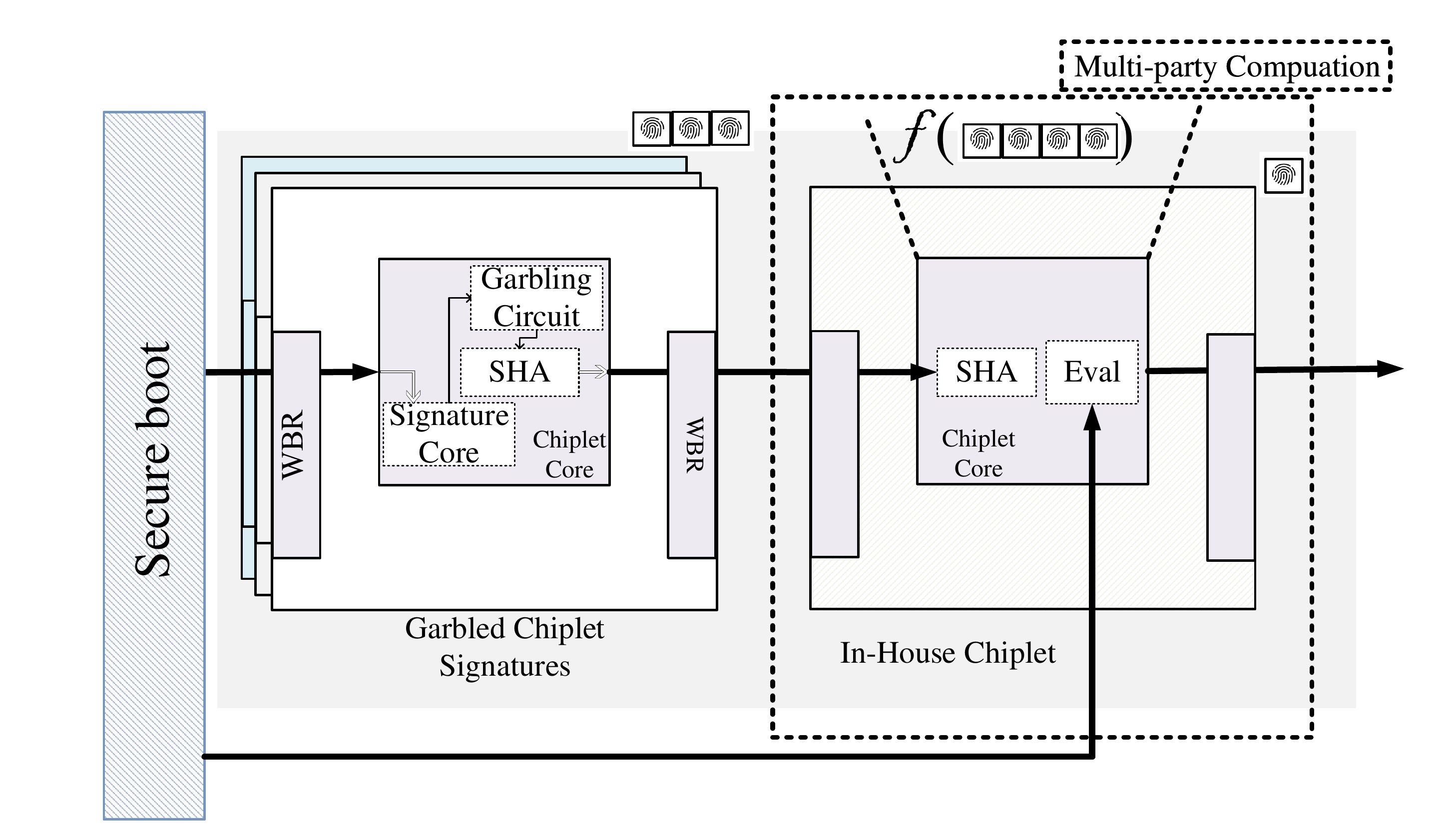}
    \caption{Detailed authentication framework using \emph{SAFE-SiP}.}
    \label{fig:framework} 
\end{figure}

\noindent\underline{\textbf{(SecObj2)}} 
Even with Garbling circuits in place for encrypted authentication, unauthorized chiplet modifications by the foundry may violate the integrity of garbled signatures ($G(S_{i})$), jeopardizing the secure MPC. A hash operation can ensure the integrity of garbled signature.  
Unauthorized chiplet modifications by the foundry cause hash mismatches, making them detectable. However, if a foundry inserts a Trojan without altering the garbled circuit or signature generation, authentication alone cannot detect it—such tampering is only identifiable through dedicated post-fabrication testing. Conversely, if an integrator and foundry collude against a vendor, the vendor’s signature remains protected through \emph{SAFE-SiP}. While authentication prevents signature leakage, hardware modifications remain undetectable unless they alter garbling or signature generation, in which case testing reveals discrepancies. 

By embedding security within core chiplet functionality instead of dedicated security IPs, \emph{SAFE-SiP} minimizes attack surfaces like Trojan insertion and signature extraction. Additionally, cryptographic transformation ensures intercepted authentication data remains infeasible to reverse-engineer, preventing unauthorized overproduction. However, \emph{SAFE-SiP} strictly verifies provenance rather than vendor trustworthiness as compromised chiplet detection is beyond authentication’s scope.

\vspace{-7pt}

\subsection{Overview of \emph{SAFE-SiP}} \label{subsec:overview}
Building on insights from Sec.~\ref{subsec:existing_techniques} and~\ref{subsec:sec_obj}, the semiconductor industry demands cost-effective security solutions that integrate seamlessly with existing signature generation methods. \emph{SAFE-SiP} addresses this need by combining a garbling circuit and a hash core (SHA-256) to enable secure and verifiable chiplet authentication. The garbling circuit obfuscates chiplet signatures, ensuring confidentiality, while the hash core generates fixed-size outputs to maintain integrity and prevent reverse engineering~\cite{gate_sip}. 
\emph{SAFE-SiP} also integrates with IEEE 1500~\cite{1500}, 1687~\cite{1687}, and 1838~\cite{1838}-compliant Design-for-Testability (DfT) components, such as the wrapper boundary register (WBR) and wrapper instruction register (WIR), enabling structured authentication without added design complexity. Moreover, while effective for digital IPs, authenticating analog IPs poses unique challenges due to their continuous nature. To bridge this gap, an Analog-to-Digital Wrapper (ADC-W) can be introduced within \emph{SAFE-SiP} to convert unique analog characteristics into secure digital signatures, extending authentication to mixed-signal systems. Hence, the chiplet vendor embeds the signature core, garbling circuit, and SHA-256, while the integrator performs verification on a trusted chiplet. By using \emph{SAFE-SiP}, existing signatures achieve:

\vspace{-3pt}
\begin{itemize}[leftmargin=*]
    \item Signature obfuscation without requiring a separate security chiplet or IP that could become an adversarial target.
    \item Protection against removal, replay, tampering, DDoS, and forging attacks while complying with IEEE 1500~\cite{ieee1500}.
    \item Data integrity and confidentiality across chiplet communications with unique, irreversible outputs in untrusted SiPs, while incurring additional low overhead in terms of area and power.
\end{itemize}
\vspace{-0.1in}

\subsection{Process Flow} \label{framework}
\begin{figure}[t] 
    \centering
    \includegraphics[width=0.45\textwidth]{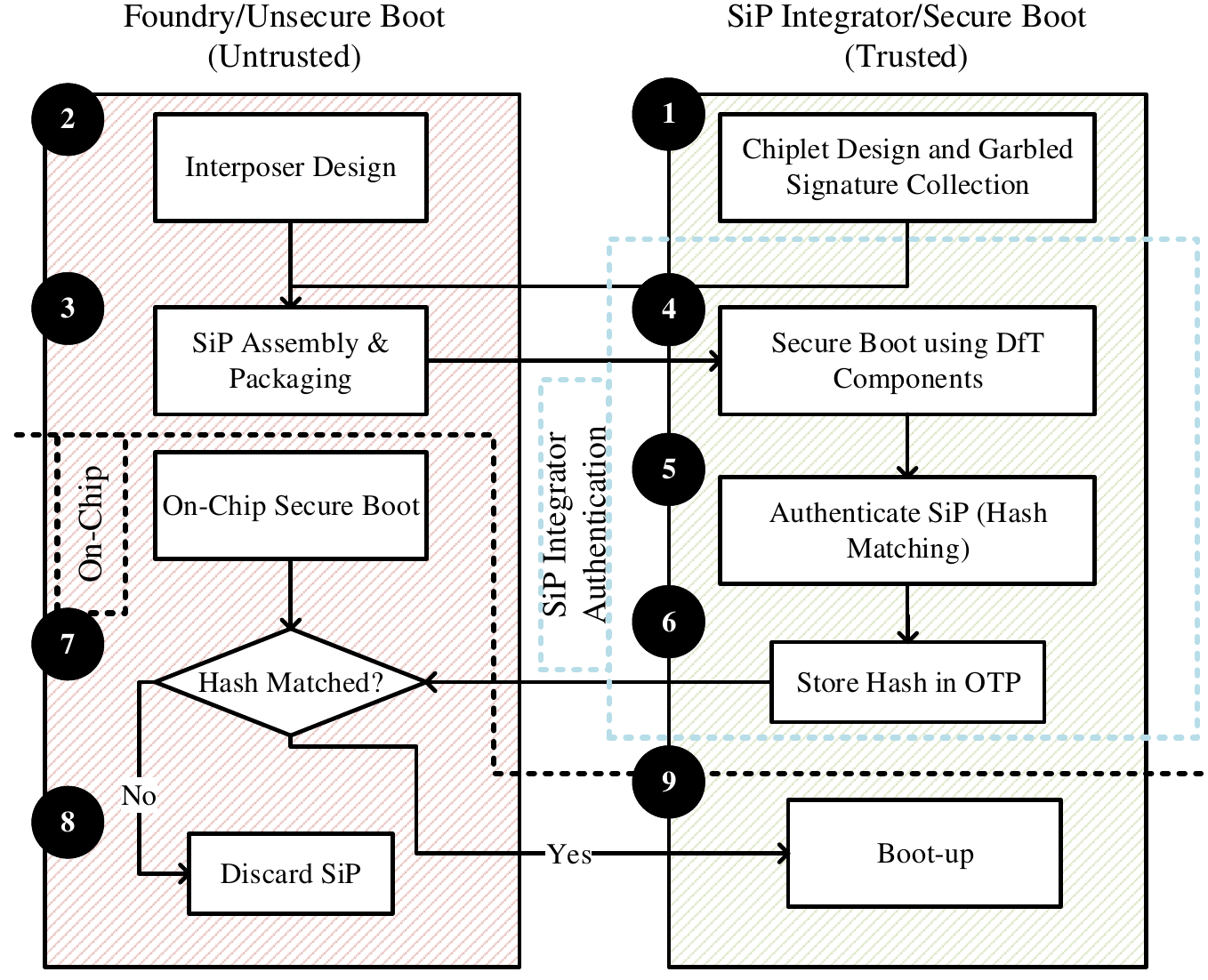}
    \caption{\emph{SAFE-SiP} process flow with supply chain and on-chip secure boot.}
    \label{fig:flow} 
\end{figure}

\textcircled{1}~The \emph{SAFE-SiP} methodology begins with the SiP integrator sourcing physical chiplets from vendors, each embedded with a watermark generation circuit capable of producing unique signatures. These signatures, provided by the vendors, act as baselines for future authentication. The garbling circuit takes these signatures as input, generates a garbled version that is provided to the SHA-256 unit for attestation before sending out of the chiplet boundary. 
\textcircled{2}~The integrator designs the interposer layout, ensuring compatibility with chiplet configurations while mitigating risks of bypass or tampering. This independent interposer design reduces vulnerabilities during the external assembly process. 
\textcircled{3}~Once chiplets and interposers are sent to a potentially untrusted packaging facility, the SiP is assembled and sent back to the SiP integrator for testing and provisioning.

\textcircled{4}~Upon return, the SiP undergoes a secure boot mechanism, leveraging the DfT components like the WBR and WIR to initiate the authentication process, following standardized IEEE protocols~\cite{1500, 1687, 1838}. The WBR facilitates direct interaction with chiplets, providing essential handshake signals and inputs for signature generation. Simultaneously, the WIR interprets control instructions to trigger authentication, ensuring seamless integration into the boot sequence.

\textcircled{5}~During authentication, each chiplet generates garbled outputs that are validated through Multi-party Computation by comparing them against vendor-provided signatures, provided prior by the vendor themselves. Any mismatch, caused by circuit alterations or tampering, generates a signature that is different that the one provided by the chiplet vendor in step $1$, leads to a different garbled and hashed output than expected, and flags the chiplet as compromised one during evaluation process by in-house chiplet. 
\textcircled{6}~Verified outputs are hashed and stored in one-time programmable (OTP) memory to facilitate future boot cycles.

\textcircled{7}~During subsequent secure boot cycles, the system reauthenticates chiplets by comparing newly generated hashes with stored values. Any mismatch disables the compromised chiplet, preventing it from impacting the system. 
\textcircled{8}~This iterative reauthentication ensures operational security, with only authenticated SiPs remaining active. 
\textcircled{9}~Following each successful secure boot, it can be ensured that the SiP is authenticated and ready. 

\textit{It is to note that, Fig.~\ref{fig:flow} illustrates the \emph{SAFE-SiP} process flow from the perspective of the SiP integrator, where the rightmost box signifies the trusted execution of the SiP integrator’s authentication and secure boot process within the supply chain, and hence, marked as trusted.}
\vspace{-6pt}

\begin{figure}[t]
    \centering
    \includegraphics[width=0.48\textwidth]{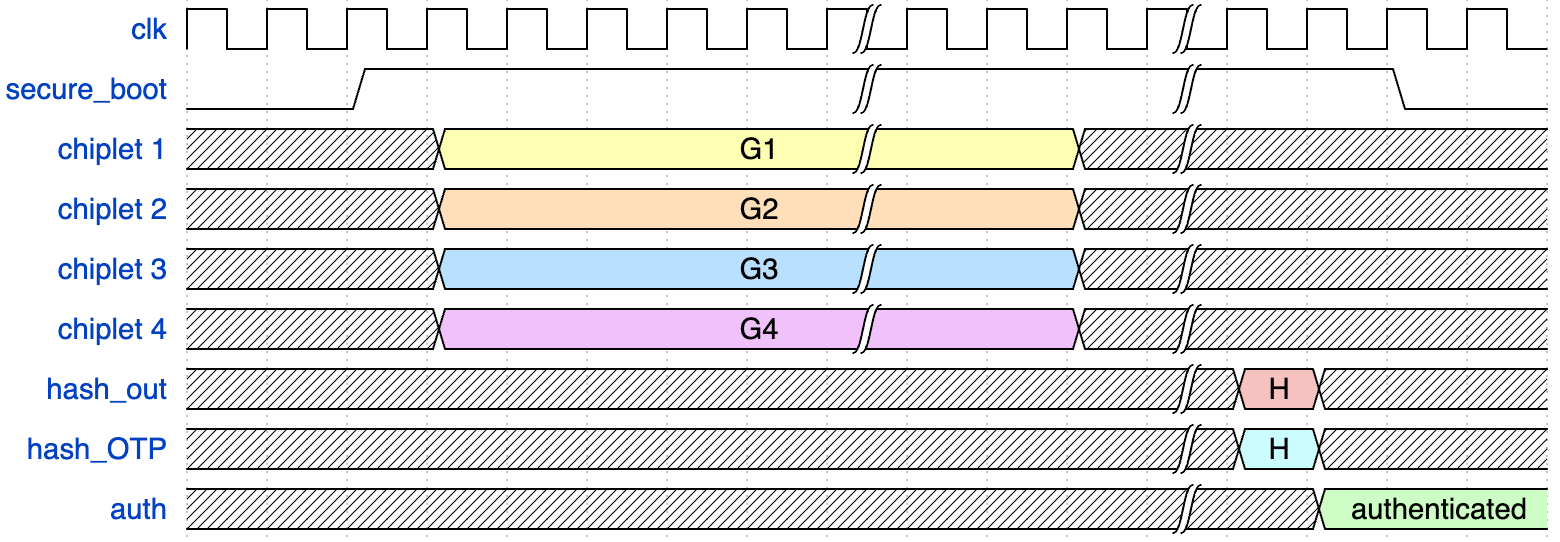}
    \caption{\emph{SAFE-SiP} waveform showcasing secure boot protocol.}
    \label{fig:waveform}
\end{figure}

\subsubsection{Garbling Circuit}\label{gc}
The \emph{SAFE-SiP} framework employs a garbling circuit to ensure the confidentiality and integrity of input signatures by transforming them into encrypted outputs. This transformation utilizes cryptographic labels and random bits generated by the chiplet’s true random number generator, enhancing unpredictability and security~\cite{gc_bmr, gc_beaver_secureprotocols}. 

The process begins by masking each bit \( b^i \) of a \( W \)-bit signature \( S \) with a random bit \( r \) and label \( L \), where one state is represented by \( r \) and its complement \( \bar{r} \), ensuring obfuscation:
\vspace{-0.075in}
\begin{align}
\tilde{b}_0^i &= r_0^i \| L_0, \\
\tilde{b}_1^i &= r_1^i \| L_1, \quad \text{where } r_1^i = \bar{r}_0^i
\end{align}
\vspace{-0.1in}

\noindent The randomness of \( r \) and \( \bar{r} \) makes each bit indistinguishable without the correct masking key. The garbling transformation is expressed:
\vspace{-0.05in}
\begin{align}
S \in (0,1)^W \rightarrow G \in (0,1)^g, \quad \text{where } g = W \cdot \kappa 
\end{align}
\vspace{-0.15in}

\noindent where \(\kappa\) is the security parameter, defining the length of labels \( r \) and \( L \). This process ensures strong encryption, preventing unauthorized data extraction and reverse engineering~\cite{rahman2020defense}. The integration of cryptographic masking and cyclic encryption secures data throughout the lifecycle, making \emph{SAFE-SiP} both robust and efficient.

\subsubsection{SHA-256 Core}
The SHA-256 hashing algorithm plays a crucial role in \emph{SAFE-SiP}, ensuring data integrity and authenticity. Recognized for its resistance to collision attacks, SHA-256 generates fixed-size hashes, making it computationally infeasible to retrieve original messages or produce identical hash outputs~\cite{crypto_engineering}. 

Within \emph{SAFE-SiP}, SHA-256 secures the garbled outputs \( G(S_i) \), verifying their integrity:
\vspace{-0.05in}
\begin{equation}
H = \text{Hash}(G(S_1), G(S_2), G(S_3), G(S_4))
\end{equation}
\vspace{-0.1in}
\begin{equation}
E = \text{Eval}(H)
\end{equation}

\noindent Any tampering with \( G(S_i) \) alters \( H \), making unauthorized modifications detectable. By leveraging SHA-256, \emph{SAFE-SiP} prevents data manipulation and ensures secure communication all over~\cite{sha256_optimizing, sha256_study}.





\vspace{-0.1in}
\begin{figure}[t] 
    \centering
    \includegraphics[width=0.48\textwidth]{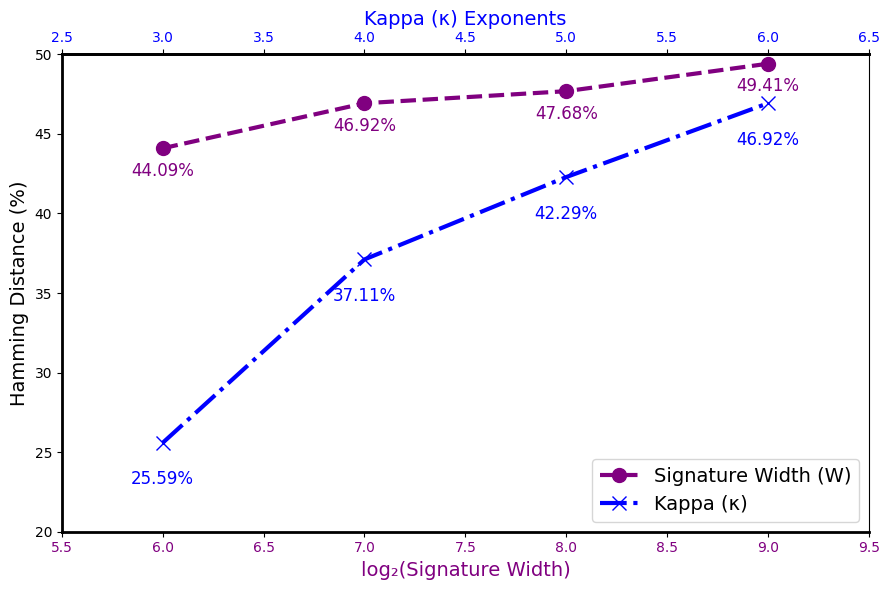}
    \caption{\emph{SAFE-SiP}'s resistance against fault injection attacks. \textit{Blue}: fixed signature width (\textit{64}), and \textit{Purple}: fixed $\kappa$ (\textit{64}).}
    \label{fig:kappavsHD} 
\end{figure}
\section{Security Analysis} \label{attack}
\emph{SAFE-SiP} utilizes built-in True Random Number Generators (TRNGs) in modern chiplets to securely generate cryptographic labels (\(L_0, L_1\)) and masking bits (\(r, \bar{r}\)) for garbling input signatures as discussed in Sec.~\ref{gc}. Unlike pseudo-random generators, TRNGs derive randomness from unpredictable physical phenomena, ensuring secure and independent parameter creation. This unpredictability is vital for the garbling process, as compromised or predictable RNGs could expose garbled outputs, enabling adversarial inferences. The following section discusses the various adversarial scenarios in SiP assembly and post-distribution and \emph{SAFE-SiP}'s resilience to them.
\vspace{-0.1in}

\subsection{Removal Attacks} \label{removal}

\noindent Removal attacks pose a significant threat in chiplet-based systems, where an adversarial foundry may capture the authentication mechanism, record its outputs, remove the underlying logic, and insert malicious modifications into the chiplet. This attack risks compromising the integrity of the chiplet and may lead to unauthorized functionality or leakage of sensitive information. The \emph{SAFE-SiP} framework is inherently resistant to removal attacks due to the integration of its garbling circuit with the watermarking circuitry of the chiplet. The garbled circuit outputs are indistinguishable from normal circuit outputs, making it extremely challenging for an attacker to identify and isolate the authentication mechanism. This obfuscation is further enhanced by the seamless embedding of the authentication process into the chiplet's operational flow, ensuring that the garbling logic is intertwined with the functional logic in a manner that does not expose a distinct authentication structure. Moreover, additional security enhancements such as logic locking and circuit camouflaging, can provide layers of protection against removal or reverse engineering. 
\vspace{-0.1in}

\subsection{Replay Attacks} \label{replay}

Replay attacks occur when adversaries reuse stored outputs or employ brute force methods~\cite{replay}. The \emph{SAFE-SiP} framework counters these attacks with layered defenses, yielding a computational complexity as show in the equation below.
\begin{equation}
TC = g \cdot 2^{64} \cdot 2^{128} = g \cdot 2^{192}
\end{equation}
\noindent Here, \(2^{64}\) corresponds to the 64-bit TRNG in garbling and \(2^{128}\) reflects the brute-force resistance of SHA-256. \(g\) denotes the signature guessing complexity determined by its width. This exponential complexity makes brute force infeasible, while the TRNG’s unique, non-deterministic outputs and SHA-256’s irreversible transformations further bolster protection against replay attacks.


\begin{table}[t]
\centering
\fontsize{7.5}{9.6}\selectfont 
\setlength\tabcolsep{5pt} 
\caption{Area Overhead Analysis of \emph{SAFE-SiP}.}
\label{tab:area_overhead}
\begin{tabular}{lcccccc}
\toprule
\textbf{\begin{tabular}[c]{@{}c@{}}Design\\Name\end{tabular}} &
\textbf{\begin{tabular}[c]{@{}c@{}}Design \\Size\\ (\( \mu m^2 \))\end{tabular}} &
\textbf{\begin{tabular}[c]{@{}c@{}}Sec.\\Param. (\( \kappa \))\\ (\#)\end{tabular}} &
\textbf{\begin{tabular}[c]{@{}c@{}}Net Area\\ Overhead \\(\%)\end{tabular}} &
\textbf{\begin{tabular}[c]{@{}c@{}}Cell Count\\ Overhead \\(\%)\end{tabular}} &
\textbf{\begin{tabular}[c]{@{}c@{}}Cell Area\\ Overhead \\(\%)\end{tabular}} \\
\midrule
\multirow{2}{*}{Ariane}   & \multirow{2}{*}{5263k} & 32 & 0.45  & 1.91  & 1.19  \\ 
                         &                         & 64 & 0.73  & 3.01  & 1.91  \\
\multirow{2}{*}{CVA6}     & \multirow{2}{*}{1271k} & 32 & 2.15  & 6.46  & 4.94  \\ 
                         &                              & 64 & 3.59  & 10.13  & 7.92  \\
\multirow{2}{*}{OR1200}   & \multirow{2}{*}{5472k} & 32 & 0.38  & 2.54  & 1.15  \\ 
                         &                              & 64 & 0.64  & 3.99  & 1.84  \\
\multirow{2}{*}{NVDLA}   & \multirow{2}{*}{1991k} & 32 & 1.29  & 5.34  & 3.15  \\ 
                         &                              & 64 & 2.16  & 8.38  & 5.05  \\
\multirow{2}{*}{RISC-V}   & \multirow{2}{*}{4815k} & 32 & 0.45  & 2.84  & 1.30  \\ 
                         &                              & 64 & 0.76  & 4.45  & 2.09  \\

\bottomrule
\end{tabular}
\end{table}

\subsection{Fault Injection or Tampering}\label{tampering}

\noindent Tampering attacks use fault injection techniques—such as power glitches or clock disruptions—to bypass chiplet authentication, posing serious risks to SoC security. \emph{SAFE-SiP} mitigates these threats by ensuring authentication fails when faults alter the process. As shown in Fig.~\ref{fig:kappavsHD}, high Hamming Distance (HD) values indicate strong fault tolerance, with disruptions causing significant output variation. The garbled circuit uses TRNG labels (\(r, L\)) and signature (\(S\)) to produce output \(G\); faults yield incorrect \(G\), with HD values reaching \(49.41\%\). SHA-256 amplifies sensitivity, where even single-bit changes in \(G\) cause hash mismatches with \(H_{\text{expected}}\). Fig.~\ref{fig:kappavsHD} shows both fixed-signature and fixed-\(\kappa\) scenarios maintain high HD with increasing size. Secure boot protocols further enhance robustness, ensuring tampered chiplets are reauthenticated and disabled. This layered defense ensures integrity under tampering.

\vspace{-0.1in}
\subsection{Denial of Service Attacks} \label{dos}

\noindent Denial attacks exploit the hierarchical structure of 2.5D and 3D chiplet-based designs, where one chiplet may block the authentication of another. In 3D integrations, this can involve a bottom chiplet denying the authentication of a top chiplet, compromising the entire stack. For 2.5D structures, the \emph{SAFE-SiP} framework addresses this by independently authenticating each chiplet through parallel testing, using its garbled signature and TRNG-generated random outputs. This ensures system-wide reliability, as no single chiplet can obstruct authentication. In more interdependent 3D stacks, if there are any man-in-the-middle (MITM) attacks, the resulting output would be different, revealing potential intruders.

\vspace{-0.1in}
\subsection{Bypass and Forging} \label{bypass} \label{forging}

\noindent The \emph{SAFE-SiP} framework ensures vigorous protection against bypass and forging attacks, safeguarding chiplet integrity and authenticity. In bypass attacks, adversaries attempt to circumvent authentication to enable unauthorized chiplets. \emph{SAFE-SiP} uses garbled circuits and SHA-256 hashing to tightly bind the chiplet's signature to its authentication output, flagging mismatches as unauthorized. Secure boot protocols further reinforce security by reauthenticating chiplets at every system boot. Forging attacks, where adversaries implant false watermarks or alter authentication, are mitigated by the randomness introduced during the garbling process, which obfuscates signature patterns and prevents replication.

\vspace{-0.1in}

\section{Design Overhead Analysis} \label{sec:result}
The following section provides an overview of \emph{SAFE-SiP}'s practicality and analyses the area, timing and power overhead.

\vspace{-0.1in}

\subsection{Experimental Setup}

We implemented \emph{SAFE-SiP} in Verilog and synthesized it using Synopsys Design Compiler using the SAED $14\,\mathrm{nm}$ standard cell library. Power, area, and timing were extracted from the post-synthesis netlist. Experiments ran on a dual-socket Intel Xeon system with $32$ physical cores, $190\,\mathrm{GB}$ RAM, based on the Skylake microarchitecture. The design was verified for functionality and synthesized to meet timing at $100\,\mathrm{MHz}$ under typical PVT conditions.

\vspace{-0.1in}

\begin{table}[t]
\centering
\fontsize{8}{9.6}\selectfont 
\setlength\tabcolsep{5pt} 
\caption{Timing Analysis for \emph{SAFE-SiP}.}
\label{tab:timing_analysis}
\begin{tabular}{cccccc}
\toprule
\textbf{\begin{tabular}[c]{@{}c@{}}Sec. Param.\\ ($\kappa$)\end{tabular}} &
\textbf{\begin{tabular}[c]{@{}c@{}}Critical Path\\ Length (ns)\end{tabular}} &
\textbf{\begin{tabular}[c]{@{}c@{}}WNS \\ (ns)\end{tabular}} &
\textbf{\begin{tabular}[c]{@{}c@{}}TNS \\ (ns)\end{tabular}} &
\textbf{\begin{tabular}[c]{@{}c@{}}Violated\\ Paths (\#)\end{tabular}} &
\textbf{\begin{tabular}[c]{@{}c@{}}Latency \\ (cc)\end{tabular}} \\
\midrule
16 & 8.50 & 0.00 & 0.00 & 0 & 96  \\
32 & 8.50 & 0.00 & 0.00 & 0 & 160 \\
64 & 8.50 & 0.00 & 0.00 & 0 & 192 \\
\bottomrule
\end{tabular}
\end{table}

\subsection{Area Overhead Analysis} \label{sec:area}
The area overhead analysis of the \emph{SAFE-SiP} framework, summarized in Table~\ref{tab:area_overhead}, underscores its efficient hardware utilization while maintaining high security. Across various designs, cell area overhead remains consistently low, reinforcing the framework’s suitability for resource-constrained environments. For a security parameter of $\kappa = 32$, OR1200 and Ariane exhibit minimal overheads of $1.15\%$ and $1.19\%$, respectively, with RISC-V following closely at $1.30\%$. NVDLA and CVA6, despite their increased complexity, maintain moderate overheads of $3.15\%$ and $4.94\%$, respectively. At the higher security level ($\kappa = 64$), the highest recorded overhead remains within $7.92\%$ (CVA6), while other designs, including OR1200 ($1.84\%$) and RISC-V ($2.09\%$), continue to demonstrate efficient area usage. These results highlight that \emph{SAFE-SiP} achieves an optimal balance between hardware efficiency and cryptographic security, ensuring chiplet authentication without significant area penalties.

\vspace{-0.1in}
\subsection{Timing Overhead Analysis} \label{sec:timing}

The \emph{SAFE-SiP} framework ensures exceptional timing integrity across varying security parameters, as summarized in Table~\ref{tab:timing_analysis}. For all evaluated configurations (\(\kappa = 16, 32, 64\)), the design consistently achieves a critical path length of $8.50~ns$ with no instances of Worst Negative Slack (WNS), Total Negative Slack (TNS), or timing violations. This demonstrates strict adherence to timing constraints, ensuring reliable operation at a reasonable scan clock frequency of $1~GHz$. The absence of violations across all configurations establishes \emph{SAFE-SiP} as a robust and predictable solution for secure computations. Additionally, the incurred authentication latency—ranging from 96 to 192 clock cycles depending on \(\kappa\)—is inherently parallelizable within different SiPs and among chiplets in a single SiP, allowing efficient distribution of authentication computations across multiple processing units. This flexibility enables seamless integration into high-performance and real-time systems, as parallel execution mitigates latency impact while maintaining security guarantees. The clean timing profile further reinforces \emph{SAFE-SiP}'s adaptability to higher clock frequencies, making it a scalable and efficient solution for secure chiplet authentication in modern SoC architectures.

\begin{table}[t]
\centering
\fontsize{7.65}{9.18}\selectfont 
\setlength\tabcolsep{12pt} 
\setlength\extrarowheight{2pt}
\caption{Power Overhead Analysis for \emph{SAFE-SiP}.}
\label{tab:power_overhead}
\begin{tabular}{lccc}
\toprule
\textbf{\begin{tabular}[c]{@{}c@{}}Design \\ Name\end{tabular}} & 
\textbf{\begin{tabular}[c]{@{}c@{}}Baseline \\ Power (mW)\end{tabular}} & 
\textbf{\begin{tabular}[c]{@{}c@{}}Sec. Param. \\ (\( \kappa \)) (\#)\end{tabular}} &
\textbf{\begin{tabular}[c]{@{}c@{}}Overhead \\ (\%)\end{tabular}} \\
\midrule
\multirow{2}{*}{Ariane} & \multirow{2}{*}{94.157} & 32 & 4.16 \\
                        &                          & 64 & 4.67 \\
\multirow{2}{*}{CVA6}   & \multirow{2}{*}{12.896} & 32 & 30.34 \\
                        &                          & 64 & 34.08 \\
\multirow{2}{*}{OR1200} & \multirow{2}{*}{106.610} & 32 & 3.67 \\
                        &                           & 64 & 4.12 \\
\multirow{2}{*}{NVDLA}  & \multirow{2}{*}{185.140} & 32 & 2.11 \\
                        &                           & 64 & 2.37 \\
\multirow{2}{*}{RISC-V} & \multirow{2}{*}{59.164}  & 32 & 6.61 \\
                        &                           & 64 & 7.43 \\
\bottomrule
\end{tabular}
\end{table}

\vspace{-0.15in}

\subsection{Power Overhead Analysis} \label{sec:power}

The power overhead analysis of the \emph{SAFE-SiP} framework, summarized in Table~\ref{tab:power_overhead}, highlights its scalability across different chiplet architectures and security parameters (\(\kappa\)). The results indicate that larger designs, such as NVDLA and OR1200, exhibit significantly lower relative power overhead compared to smaller designs like CVA6. For instance, NVDLA incurs only $2.37\%$ and $2.11\%$ overhead for \(\kappa = 64\) and \(32\), respectively, while OR1200 maintains similarly low values of $4.12\%$ and $3.67\%$. This trend underscores \emph{SAFE-SiP}'s efficiency in leveraging higher baseline power in larger designs, minimizing the relative computational overhead. Conversely, in the smaller CVA6 design, the power overhead is more pronounced, reaching $34.08\%$ and $30.34\%$ for the same security parameters due to its lower baseline power, yet it remains within practical limits for resource-constrained deployments. RISC-V demonstrates a balanced efficiency, with overhead values of $7.43\%$ and $6.61\%$, further reinforcing the framework’s adaptability. These findings confirm that \emph{SAFE-SiP} achieves better energy efficiency as chiplet size increases, making it an optimal choice for integrating secure authentication in high-performance architectures while maintaining reasonable overhead for smaller systems.

\vspace{-0.1in}

\section{Conclusion}
\label{sec:conclusion}

This work presents \emph{SAFE-SiP}, a scalable authentication framework for 2.5D/3D SiP assemblies using Multi-party Computation and garbled circuits. By garbling and hashing chiplet signatures, \emph{SAFE-SiP} ensures confidentiality for vendors and detects tampering by foundries. It integrates with diverse signature schemes and secure boot processes without requiring dedicated security hardware. Security analysis confirms resilience to spoofing, tampering, and replay attacks using SHA-256 and garbling. Evaluations on five benchmark designs show minimal area, power, and latency overhead. The implementation is open-sourced to promote collaboration.
\vspace{-6pt}
\bibliographystyle{ACM-Reference-Format}
\bibliography{sample-base}


\begin{thebibliography}{35}


\ifx \showCODEN    \undefined \def \showCODEN     #1{\unskip}     \fi
\ifx \showISBNx    \undefined \def \showISBNx     #1{\unskip}     \fi
\ifx \showISBNxiii \undefined \def \showISBNxiii  #1{\unskip}     \fi
\ifx \showISSN     \undefined \def \showISSN      #1{\unskip}     \fi
\ifx \showLCCN     \undefined \def \showLCCN      #1{\unskip}     \fi
\ifx \shownote     \undefined \def \shownote      #1{#1}          \fi
\ifx \showarticletitle \undefined \def \showarticletitle #1{#1}   \fi
\ifx \showURL      \undefined \def \showURL       {\relax}        \fi
\providecommand\bibfield[2]{#2}
\providecommand\bibinfo[2]{#2}
\providecommand\natexlab[1]{#1}
\providecommand\showeprint[2][]{arXiv:#2}

\bibitem[iee(2022)]%
        {ieee1500}
 \bibinfo{year}{2022}\natexlab{}.
\newblock \showarticletitle{IEEE Standard Testability Method for Embedded Core-based Integrated Circuits}.
\newblock \bibinfo{journal}{\emph{IEEE Std 1500-2022 (Revision of IEEE Std 1500-2005)}} (\bibinfo{year}{2022}), \bibinfo{pages}{1--168}.
\newblock
\href{https://doi.org/10.1109/IEEESTD.2022.9916221}{doi:\nolinkurl{10.1109/IEEESTD.2022.9916221}}


\bibitem[Ahmed~et al.(2024)]%
        {ll_seamless}
\bibfield{author}{\bibinfo{person}{Bulbul Ahmed~et al.}} \bibinfo{year}{2024}\natexlab{}.
\newblock \showarticletitle{SeeMLess: Security Evaluation of Logic Locking using Machine Learning oriented Estimation}. In \bibinfo{booktitle}{\emph{Proceedings of the Great Lakes Symposium on VLSI 2024}} (Clearwater, FL, USA) \emph{(\bibinfo{series}{GLSVLSI '24})}. \bibinfo{publisher}{Association for Computing Machinery}, \bibinfo{address}{New York, NY, USA}, \bibinfo{pages}{489–494}.
\newblock
\showISBNx{9798400706059}
\href{https://doi.org/10.1145/3649476.3660382}{doi:\nolinkurl{10.1145/3649476.3660382}}


\bibitem[Alkabani et~al\mbox{.}(2007)]%
        {alkabani2007active}
\bibfield{author}{\bibinfo{person}{Yousra Alkabani}, \bibinfo{person}{Farinaz Koushanfar}, {et~al\mbox{.}}} \bibinfo{year}{2007}\natexlab{}.
\newblock \showarticletitle{Active Hardware Metering for Intellectual Property Protection and Security.}. In \bibinfo{booktitle}{\emph{USENIX security symposium}}, Vol.~\bibinfo{volume}{20}. \bibinfo{pages}{1--20}.
\newblock


\bibitem[Anandakumar et~al\mbox{.}(2022)]%
        {anandakumar2022}
\bibfield{author}{\bibinfo{person}{N.~Nalla Anandakumar} {et~al\mbox{.}}} \bibinfo{year}{2022}\natexlab{}.
\newblock \showarticletitle{Rethinking Watermark: Providing Proof of {IP} Ownership in Modern SoCs}.
\newblock \bibinfo{journal}{\emph{{IACR} Cryptol. ePrint Arch.}} (\bibinfo{year}{2022}).
\newblock


\bibitem[Beaver et~al\mbox{.}(1990)]%
        {gc_beaver_secureprotocols}
\bibfield{author}{\bibinfo{person}{D. Beaver}, \bibinfo{person}{S. Micali}, {and} \bibinfo{person}{P. Rogaway}.} \bibinfo{year}{1990}\natexlab{}.
\newblock \showarticletitle{The round complexity of secure protocols}. In \bibinfo{booktitle}{\emph{Proceedings of the Twenty-Second Annual ACM Symposium on Theory of Computing}} (Baltimore, Maryland, USA) \emph{(\bibinfo{series}{STOC '90})}. \bibinfo{publisher}{Association for Computing Machinery}, \bibinfo{address}{New York, NY, USA}, \bibinfo{pages}{503–513}.
\newblock
\showISBNx{0897913612}
\href{https://doi.org/10.1145/100216.100287}{doi:\nolinkurl{10.1145/100216.100287}}


\bibitem[Bellare et~al\mbox{.}(2013)]%
        {gc_fkb}
\bibfield{author}{\bibinfo{person}{Mihir Bellare}, \bibinfo{person}{Viet~Tung Hoang}, \bibinfo{person}{Sriram Keelveedhi}, {and} \bibinfo{person}{Phillip Rogaway}.} \bibinfo{year}{2013}\natexlab{}.
\newblock \showarticletitle{Efficient Garbling from a Fixed-Key Blockcipher}. In \bibinfo{booktitle}{\emph{2013 IEEE Symposium on Security and Privacy}}. \bibinfo{pages}{478--492}.
\newblock
\href{https://doi.org/10.1109/SP.2013.39}{doi:\nolinkurl{10.1109/SP.2013.39}}


\bibitem[Ben-David et~al\mbox{.}(2008)]%
        {gc_fairplay}
\bibfield{author}{\bibinfo{person}{Assaf Ben-David}, \bibinfo{person}{Noam Nisan}, {and} \bibinfo{person}{Benny Pinkas}.} \bibinfo{year}{2008}\natexlab{}.
\newblock \showarticletitle{FairplayMP: a system for secure multi-party computation}. In \bibinfo{booktitle}{\emph{Proceedings of the 15th ACM Conference on Computer and Communications Security}} (Alexandria, Virginia, USA) \emph{(\bibinfo{series}{CCS '08})}. \bibinfo{publisher}{Association for Computing Machinery}, \bibinfo{address}{New York, NY, USA}, \bibinfo{pages}{257–266}.
\newblock
\showISBNx{9781595938107}
\href{https://doi.org/10.1145/1455770.1455804}{doi:\nolinkurl{10.1145/1455770.1455804}}


\bibitem[Chandra~et al.(2023)]%
        {ieee1838_casestudy}
\bibfield{author}{\bibinfo{person}{Anshuman Chandra~et al.}} \bibinfo{year}{2023}\natexlab{}.
\newblock \showarticletitle{A Case Study on IEEE 1838 Compliant Multi-Die 3DIC DFT Implementation}. In \bibinfo{booktitle}{\emph{2023 IEEE International Test Conference (ITC)}}. \bibinfo{pages}{11--20}.
\newblock
\href{https://doi.org/10.1109/ITC51656.2023.00011}{doi:\nolinkurl{10.1109/ITC51656.2023.00011}}


\bibitem[Chen~et al.(2019)]%
        {3d_soictm}
\bibfield{author}{\bibinfo{person}{Ming-Fa Chen~et al.}} \bibinfo{year}{2019}\natexlab{}.
\newblock \showarticletitle{System on Integrated Chips (SoIC(TM) for 3D Heterogeneous Integration}. In \bibinfo{booktitle}{\emph{2019 IEEE 69th Electronic Components and Technology Conference (ECTC)}}. \bibinfo{pages}{594--599}.
\newblock
\href{https://doi.org/10.1109/ECTC.2019.00095}{doi:\nolinkurl{10.1109/ECTC.2019.00095}}


\bibitem[Courtois~et al.(2014)]%
        {sha256_optimizing}
\bibfield{author}{\bibinfo{person}{Nicolas~T. Courtois~et al.}} \bibinfo{year}{2014}\natexlab{}.
\newblock \showarticletitle{Optimizing SHA256 in Bitcoin Mining}. In \bibinfo{booktitle}{\emph{Cryptography and Security Systems}}, \bibfield{editor}{\bibinfo{person}{Zbigniew Kotulski}, \bibinfo{person}{Bogdan Ksi{\k{e}}{\.{z}}opolski}, {and} \bibinfo{person}{Katarzyna Mazur}} (Eds.). \bibinfo{publisher}{Springer Berlin Heidelberg}, \bibinfo{address}{Berlin, Heidelberg}, \bibinfo{pages}{131--144}.
\newblock
\showISBNx{978-3-662-44893-9}


\bibitem[Deric and Holcomb(2022)]%
        {deric2022know}
\bibfield{author}{\bibinfo{person}{Aleksa Deric} {and} \bibinfo{person}{Daniel Holcomb}.} \bibinfo{year}{2022}\natexlab{}.
\newblock \showarticletitle{Know time to die--integrity checking for zero trust chiplet-based systems using between-die delay PUFs}.
\newblock \bibinfo{journal}{\emph{IACR Transactions on Cryptographic Hardware and Embedded Systems}} (\bibinfo{year}{2022}), \bibinfo{pages}{391--412}.
\newblock


\bibitem[et~al.(2022)]%
        {toshi}
\bibfield{author}{\bibinfo{person}{Nidish~Vashistha et al.}} \bibinfo{year}{2022}\natexlab{}.
\newblock \bibinfo{title}{{ToSHI} - Towards Secure Heterogeneous Integration: Security Risks, Threat Assessment, and Assurance}.
\newblock \bibinfo{howpublished}{Cryptology {ePrint} Archive, Paper 2022/984}.
\newblock
\urldef\tempurl%
\url{https://eprint.iacr.org/2022/984}
\showURL{%
\tempurl}


\bibitem[Farheen~et al.(2023)]%
        {twofold_laser}
\bibfield{author}{\bibinfo{person}{Tasnuva Farheen~et al.}} \bibinfo{year}{2023}\natexlab{}.
\newblock \showarticletitle{A Twofold Clock and Voltage-Based Detection Method for Laser Logic State Imaging Attack}.
\newblock \bibinfo{journal}{\emph{IEEE Transactions on Very Large Scale Integration (VLSI) Systems}} \bibinfo{volume}{31}, \bibinfo{number}{1} (\bibinfo{year}{2023}), \bibinfo{pages}{65--78}.
\newblock
\href{https://doi.org/10.1109/TVLSI.2022.3214724}{doi:\nolinkurl{10.1109/TVLSI.2022.3214724}}


\bibitem[Ferguson et~al\mbox{.}(2010)]%
        {crypto_engineering}
\bibfield{author}{\bibinfo{person}{Niels Ferguson}, \bibinfo{person}{Bruce Schneier}, {and} \bibinfo{person}{Tadayoshi Kohno}.} \bibinfo{year}{2010}\natexlab{}.
\newblock \bibinfo{booktitle}{\emph{Cryptography Engineering: Design Principles and Practical Applications}}.
\newblock \bibinfo{publisher}{Wiley Publishing}.
\newblock
\showISBNx{0470474246}


\bibitem[Haidar~et al.({[n.\,d.]})]%
        {sect_hi}
\bibfield{author}{\bibinfo{person}{Galib~Ibne Haidar~et al.}} \bibinfo{year}{[n.\,d.]}\natexlab{}.
\newblock \showarticletitle{SECT-HI: Enabling Secure Testing for Heterogeneous Integration to Prevent SiP Counterfeits}.
\newblock  (\bibinfo{year}{[n.\,d.]}).
\newblock


\bibitem[Haidar~et al(2024)]%
        {gate_sip}
\bibfield{author}{\bibinfo{person}{Galib~Ibne Haidar~et al}.} \bibinfo{year}{2024}\natexlab{}.
\newblock \showarticletitle{GATE-SiP: Enabling Authenticated Encryption Testing in Systems-in-Package}. In \bibinfo{booktitle}{\emph{Proceedings of the 61st ACM/IEEE Design Automation Conference}} (San Francisco, CA, USA) \emph{(\bibinfo{series}{DAC '24})}. \bibinfo{publisher}{Association for Computing Machinery}, \bibinfo{address}{New York, NY, USA}, Article \bibinfo{articleno}{299}, \bibinfo{numpages}{6}~pages.
\newblock
\showISBNx{9798400706011}
\href{https://doi.org/10.1145/3649329.3656527}{doi:\nolinkurl{10.1145/3649329.3656527}}


\bibitem[Haque~et al.(2023)]%
        {haque2023shi}
\bibfield{author}{\bibinfo{person}{Md~Saad~Ul Haque~et al.}} \bibinfo{year}{2023}\natexlab{}.
\newblock \showarticletitle{SHI-Lock: Enabling Co-Obfuscation for Secure Heterogeneous Integration Against RE and Cloning}. In \bibinfo{booktitle}{\emph{2023 IEEE Physical Assurance and Inspection of Electronics (PAINE)}}. IEEE, \bibinfo{pages}{1--7}.
\newblock


\bibitem[Hazay et~al\mbox{.}(2020)]%
        {gc_bmr}
\bibfield{author}{\bibinfo{person}{Carmit Hazay}, \bibinfo{person}{Peter Scholl}, {and} \bibinfo{person}{Eduardo Soria-Vazquez}.} \bibinfo{year}{2020}\natexlab{}.
\newblock \showarticletitle{Low Cost Constant Round MPC Combining BMR and Oblivious Transfer}.
\newblock \bibinfo{journal}{\emph{Journal of Cryptology}} \bibinfo{volume}{33}, \bibinfo{number}{4} (\bibinfo{date}{01 Oct} \bibinfo{year}{2020}), \bibinfo{pages}{1732--1786}.
\newblock
\showISSN{1432-1378}
\href{https://doi.org/10.1007/s00145-020-09355-y}{doi:\nolinkurl{10.1007/s00145-020-09355-y}}


\bibitem[Ibnat~et al.(2023)]%
        {watermark_actiwate}
\bibfield{author}{\bibinfo{person}{Zahin Ibnat~et al.}} \bibinfo{year}{2023}\natexlab{}.
\newblock \showarticletitle{ActiWate: Adaptive and Design-agnostic Active Watermarking for IP Ownership in Modern SoCs}. In \bibinfo{booktitle}{\emph{2023 60th ACM/IEEE Design Automation Conference (DAC)}}. \bibinfo{pages}{1--6}.
\newblock
\href{https://doi.org/10.1109/DAC56929.2023.10247688}{doi:\nolinkurl{10.1109/DAC56929.2023.10247688}}


\bibitem[Larson et~al\mbox{.}(2015)]%
        {gc_secure_auction}
\bibfield{author}{\bibinfo{person}{Maya Larson}, \bibinfo{person}{Chunqiang Hu}, \bibinfo{person}{Ruinian Li}, \bibinfo{person}{Wei Li}, {and} \bibinfo{person}{Xiuzhen Cheng}.} \bibinfo{year}{2015}\natexlab{}.
\newblock \showarticletitle{Secure Auctions without an Auctioneer via Verifiable Secret Sharing}. In \bibinfo{booktitle}{\emph{Proceedings of the 2015 Workshop on Privacy-Aware Mobile Computing}} (Hangzhou, China) \emph{(\bibinfo{series}{PAMCO '15})}. \bibinfo{publisher}{Association for Computing Machinery}, \bibinfo{address}{New York, NY, USA}, \bibinfo{pages}{1–6}.
\newblock
\showISBNx{9781450335232}
\href{https://doi.org/10.1145/2757302.2757305}{doi:\nolinkurl{10.1145/2757302.2757305}}


\bibitem[Pavlidis and Friedman(2008)]%
        {3dic_book_pavlidis}
\bibfield{author}{\bibinfo{person}{Vasilis~F. Pavlidis} {and} \bibinfo{person}{Eby~G. Friedman}.} \bibinfo{year}{2008}\natexlab{}.
\newblock \bibinfo{booktitle}{\emph{Three-dimensional Integrated Circuit Design}}.
\newblock \bibinfo{publisher}{Morgan Kaufmann Publishers Inc.}, \bibinfo{address}{San Francisco, CA, USA}.
\newblock
\showISBNx{9780080921860}


\bibitem[Rachmawati et~al\mbox{.}(2018)]%
        {sha256_study}
\bibfield{author}{\bibinfo{person}{D Rachmawati}, \bibinfo{person}{J~T Tarigan}, {and} \bibinfo{person}{A~B~C Ginting}.} \bibinfo{year}{2018}\natexlab{}.
\newblock \showarticletitle{A comparative study of Message Digest 5(MD5) and SHA256 algorithm}.
\newblock \bibinfo{journal}{\emph{Journal of Physics: Conference Series}} \bibinfo{volume}{978}, \bibinfo{number}{1} (\bibinfo{date}{mar} \bibinfo{year}{2018}), \bibinfo{pages}{012116}.
\newblock
\href{https://doi.org/10.1088/1742-6596/978/1/012116}{doi:\nolinkurl{10.1088/1742-6596/978/1/012116}}


\bibitem[Rahman~et al.(2020)]%
        {rahman2020defense}
\bibfield{author}{\bibinfo{person}{M~Tanjidur Rahman~et al.}} \bibinfo{year}{2020}\natexlab{}.
\newblock \showarticletitle{Defense-in-depth: A recipe for logic locking to prevail}.
\newblock \bibinfo{journal}{\emph{Integration}}  \bibinfo{volume}{72} (\bibinfo{year}{2020}), \bibinfo{pages}{39--57}.
\newblock


\bibitem[Rahman~et al(2023)]%
        {postfab_lle}
\bibfield{author}{\bibinfo{person}{Sazadur Rahman~et al}.} \bibinfo{year}{2023}\natexlab{}.
\newblock \showarticletitle{Lle: Mitigating ic piracy and reverse engineering by last level edit}. In \bibinfo{booktitle}{\emph{International Symposium for Testing and Failure Analysis}}, Vol.~\bibinfo{volume}{84741}. ASM International, \bibinfo{pages}{360--369}.
\newblock


\bibitem[Riazi et~al\mbox{.}(2019)]%
        {mpcircuits}
\bibfield{author}{\bibinfo{person}{M.~Sadegh Riazi}, \bibinfo{person}{Mojan Javaheripi}, \bibinfo{person}{Siam~U. Hussain}, {and} \bibinfo{person}{Farinaz Koushanfar}.} \bibinfo{year}{2019}\natexlab{}.
\newblock \showarticletitle{MPCircuits: Optimized Circuit Generation for Secure Multi-Party Computation}. In \bibinfo{booktitle}{\emph{2019 IEEE International Symposium on Hardware Oriented Security and Trust (HOST)}}. \bibinfo{pages}{198--207}.
\newblock
\href{https://doi.org/10.1109/HST.2019.8740831}{doi:\nolinkurl{10.1109/HST.2019.8740831}}


\bibitem[SA(2024a)]%
        {1687}
\bibfield{author}{\bibinfo{person}{IEEE SA}.} \bibinfo{year}{2024}\natexlab{a}.
\newblock \bibinfo{title}{IEEE Standard for Access and Control of Instrumentation Embedded within a Semiconductor Device}.
\newblock
\urldef\tempurl%
\url{https://standards.ieee.org/ieee/1687/3931/}
\showURL{%
\tempurl}
\newblock
\shownote{Accessed: November 2024}.


\bibitem[SA(2024b)]%
        {1838}
\bibfield{author}{\bibinfo{person}{IEEE SA}.} \bibinfo{year}{2024}\natexlab{b}.
\newblock \bibinfo{title}{IEEE Standard for Test Access Architecture for Three-Dimensional Stacked Integrated Circuits}.
\newblock
\urldef\tempurl%
\url{https://standards.ieee.org/ieee/1838/5073/}
\showURL{%
\tempurl}
\newblock
\shownote{Accessed: November 2024}.


\bibitem[SA(2024c)]%
        {1500}
\bibfield{author}{\bibinfo{person}{IEEE SA}.} \bibinfo{year}{2024}\natexlab{c}.
\newblock \bibinfo{title}{IEEE Standard Testability Method for Embedded Core-based Integrated Circuits}.
\newblock
\urldef\tempurl%
\url{https://standards.ieee.org/ieee/1500/7704/}
\showURL{%
\tempurl}
\newblock
\shownote{Accessed: November 2024}.


\bibitem[Safari et~al\mbox{.}(2023)]%
        {splitcore}
\bibfield{author}{\bibinfo{person}{Yousef Safari}, \bibinfo{person}{Pooya Aghanoury}, \bibinfo{person}{Subramanian~S. Iyer}, \bibinfo{person}{Nader Sehatbakhsh}, {and} \bibinfo{person}{Boris Vaisband}.} \bibinfo{year}{2023}\natexlab{}.
\newblock \showarticletitle{Hybrid Obfuscation of Chiplet-Based Systems}. In \bibinfo{booktitle}{\emph{2023 60th ACM/IEEE Design Automation Conference (DAC)}}. \bibinfo{pages}{1--6}.
\newblock
\href{https://doi.org/10.1109/DAC56929.2023.10247738}{doi:\nolinkurl{10.1109/DAC56929.2023.10247738}}


\bibitem[Sheikh~et al.(2021)]%
        {2.5_3D_hetero}
\bibfield{author}{\bibinfo{person}{Farhana Sheikh~et al.}} \bibinfo{year}{2021}\natexlab{}.
\newblock \showarticletitle{2.5D and 3D Heterogeneous Integration: Emerging applications}.
\newblock \bibinfo{journal}{\emph{IEEE Solid-State Circuits Magazine}} \bibinfo{volume}{13}, \bibinfo{number}{4} (\bibinfo{year}{2021}), \bibinfo{pages}{77--87}.
\newblock
\href{https://doi.org/10.1109/MSSC.2021.3111386}{doi:\nolinkurl{10.1109/MSSC.2021.3111386}}


\bibitem[Songhori et~al\mbox{.}(2015)]%
        {tinygarble}
\bibfield{author}{\bibinfo{person}{Ebrahim~M. Songhori}, \bibinfo{person}{Siam~U. Hussain}, \bibinfo{person}{Ahmad-Reza Sadeghi}, \bibinfo{person}{Thomas Schneider}, {and} \bibinfo{person}{Farinaz Koushanfar}.} \bibinfo{year}{2015}\natexlab{}.
\newblock \showarticletitle{TinyGarble: Highly Compressed and Scalable Sequential Garbled Circuits}. In \bibinfo{booktitle}{\emph{2015 IEEE Symposium on Security and Privacy}}. \bibinfo{pages}{411--428}.
\newblock
\href{https://doi.org/10.1109/SP.2015.32}{doi:\nolinkurl{10.1109/SP.2015.32}}


\bibitem[Taheri~et al.(2024)]%
        {script_dos}
\bibfield{author}{\bibinfo{person}{Ebadollah Taheri~et al.}} \bibinfo{year}{2024}\natexlab{}.
\newblock \showarticletitle{SCRIPT: A Multi-Objective Routing Framework for Securing Chiplet Systems against Distributed DoS Attacks}. In \bibinfo{booktitle}{\emph{Proceedings of the Great Lakes Symposium on VLSI 2024}} (Clearwater, FL, USA) \emph{(\bibinfo{series}{GLSVLSI '24})}. \bibinfo{publisher}{Association for Computing Machinery}, \bibinfo{address}{New York, NY, USA}, \bibinfo{pages}{78–85}.
\newblock
\showISBNx{9798400706059}
\href{https://doi.org/10.1145/3649476.3658763}{doi:\nolinkurl{10.1145/3649476.3658763}}


\bibitem[Ul~Islam Sami~et al.(2024)]%
        {pqc_hi}
\bibfield{author}{\bibinfo{person}{Md~Sami Ul~Islam Sami~et al.}} \bibinfo{year}{2024}\natexlab{}.
\newblock \showarticletitle{PQC-HI: PQC-enabled Chiplet Authentication and Key Exchange in Heterogeneous Integration}. In \bibinfo{booktitle}{\emph{2024 IEEE 74th Electronic Components and Technology Conference (ECTC)}}. \bibinfo{pages}{464--471}.
\newblock
\href{https://doi.org/10.1109/ECTC51529.2024.00079}{doi:\nolinkurl{10.1109/ECTC51529.2024.00079}}


\bibitem[Ustun et~al\mbox{.}(2019)]%
        {replay}
\bibfield{author}{\bibinfo{person}{Taha~Selim Ustun}, \bibinfo{person}{Shaik~Mullapathi Farooq}, {and} \bibinfo{person}{S.~M.~Suhail Hussain}.} \bibinfo{year}{2019}\natexlab{}.
\newblock \showarticletitle{A Novel Approach for Mitigation of Replay and Masquerade Attacks in Smartgrids Using IEC 61850 Standard}.
\newblock \bibinfo{journal}{\emph{IEEE Access}}  \bibinfo{volume}{7} (\bibinfo{year}{2019}), \bibinfo{pages}{156044--156053}.
\newblock
\href{https://doi.org/10.1109/ACCESS.2019.2948117}{doi:\nolinkurl{10.1109/ACCESS.2019.2948117}}


\bibitem[Yao(1986)]%
        {gc_yao}
\bibfield{author}{\bibinfo{person}{Andrew Chi-Chih Yao}.} \bibinfo{year}{1986}\natexlab{}.
\newblock \showarticletitle{How to generate and exchange secrets}. In \bibinfo{booktitle}{\emph{27th Annual Symposium on Foundations of Computer Science (sfcs 1986)}}. \bibinfo{pages}{162--167}.
\newblock
\href{https://doi.org/10.1109/SFCS.1986.25}{doi:\nolinkurl{10.1109/SFCS.1986.25}}


\end{thebibliography}

\end{document}